\documentclass[conference]{IEEEtran}
\IEEEoverridecommandlockouts
\usepackage{cite}
\usepackage{amsmath,amssymb,amsfonts}
\usepackage{algorithmic}
\usepackage{graphicx}
\usepackage{textcomp}
\usepackage{xcolor}

\usepackage{tikz}
\tikzset{near start abs/.style={xshift=1cm}}
\usetikzlibrary{plotmarks, shapes, positioning, arrows, decorations.markings, fit, calc, patterns, circuits.logic.US}
\usetikzlibrary{arrows.meta}
\usetikzlibrary{decorations.pathmorphing}
\usepackage{pgfplots} 
\usepackage{threeparttable}
\usepackage{glossaries} 
\usepackage[draft,pdfborder={0 0 0}]{hyperref}

\usepackage{multirow}
\usepackage{flushend} 

\def\BibTeX{{\rm B\kern-.05em{\sc i\kern-.025em b}\kern-.08em
    T\kern-.1667em\lower.7ex\hbox{E}\kern-.125emX}}

\newcommand{\gear}[6]{%
  (0:#2)
  \foreach \i [evaluate=\i as \n using {\i-1)*360/#1}] in {1,...,#1}{%
    arc (\n:\n+#4:#2) {[rounded corners=1.5pt] -- (\n+#4+#5:#3)
    arc (\n+#4+#5:\n+360/#1-#5:#3)} --  (\n+360/#1:#2)
  }%
  (0,0) circle[radius=#6] 
}
\usepackage{fancyhdr}
\usepackage{lipsum} 

\fancypagestyle{plain}{%
  \fancyhf{} 
}

\begin{document}

\tikzstyle{line:thick}=[line width=0.5mm, arrows = {-Latex[width=5pt, length=5pt]}]
\tikzstyle{line:thickAr}=[line width=0.5mm, arrows = {Latex[width=5pt, length=5pt]-Latex[width=5pt, length=5pt]}]

\newacronym{asic}{ASIC}{application specific integrated circuit}
\newacronym{addr}{ADR}{address}
\newacronym{alu}{ALU}{arithmetic logic unit}
\newacronym{bpu}{BPU}{blockchain processing unit}
\newacronym{bcd}{BCM}{bytecode memory}
\newacronym{cc}{CC}{clock cycle}
\newacronym{cpu}{CPU}{central processing unit}
\newacronym{dapp}{dApp}{decentralized application}
\newacronym{defi}{DeFi}{decentralized finance}
\newacronym{dma}{DMA}{direct memory access}
\newacronym{dos}{DoS}{denial of service}
\newacronym{dsp}{DSP}{digital signal processor}
\newacronym{eth}{ETH}{Ether}
\newacronym{evm}{EVM}{Ethereum virtual machine}
\newacronym{fpga}{FPGA}{field programmable gate array}
\newacronym{fsm}{FSM}{finite state machine}
\newacronym{gs}{GS}{gas}
\newacronym{hw}{HW}{hardware}
\newacronym{isa}{ISA}{instruction set architecture}
\newacronym{kecc}{KEC}{Keccak256}
\newacronym{lifo}{LIFO}{last-in, first-out}
\newacronym{lut}{LUT}{look up table}
\newacronym{lut}{LUT}{look-up table}
\newacronym{mem}{MEM}{memory}
\newacronym{op}{opcode}{operational code}
\newacronym{pc}{PC}{program counter}
\newacronym{pos}{PoS}{Proof-of-Stake}
\newacronym{pow}{PoW}{Proof-of Work}
\newacronym{ps}{PS}{processing system}
\newacronym{pl}{PL}{programmable logic}
\newacronym{ram}{RAM}{random access memory}
\newacronym{rpl}{RPL}{recursive length prefix}
\newacronym{retn}{RTN}{return memory}
\newacronym{sc}{SC}{smart contract}
\newacronym{scu}{SCU}{smart contract unit}
\newacronym{stk}{STK}{stack}
\newacronym{str}{STR}{storage}
\newacronym{sw}{SW}{software}
\newacronym{tps}{TPS}{transactions per second}

\title{EVMx: An FPGA-Based Smart Contract Processing Unit\\ 
\thanks{
2836-3795/25/\$31.00 ©2025 IEEE\newline
DOI 10.1109/COMPSAC65507.2025.00231
}
}

\author{Joel Poncha Lemayian\textsuperscript{*}, Hachem Bensalem\textsuperscript{*}, Ghyslain Gagnon\textsuperscript{*}, Kaiwen Zhang\textsuperscript{†}, and Pascal Giard\textsuperscript{*} \\ 
\textsuperscript{*}Department of Electrical Engineering, \'{E}cole de technologie sup\'{e}rieure (\'{E}TS), Montr\'{e}al, Canada\\
\textsuperscript{†}Department of Software Engineering and IT, \'{E}cole de technologie sup\'{e}rieure (\'{E}TS), Montr\'{e}al, Canada\\
Email: joel-poncha.lemayian.1@ens.etsmtl.ca, \{hachem.bensalem, kaiwen.zhang,  pascal.giard\}@etsmtl.ca}

\maketitle
\thispagestyle{fancy} 
\begin{abstract}

Ethereum blockchain uses smart contracts (SCs) to implement decentralized applications (dApps). SCs are executed by the Ethereum virtual machine (EVM) running within an Ethereum client. Moreover, the EVM has been widely adopted by other blockchain platforms, including Solana, Cardano, Avalanche, Polkadot, and more. However, the EVM performance is limited by the constraints of the general-purpose computer it operates on. This work proposes offloading SC execution onto a dedicated hardware-based EVM. Specifically, EVMx is an FPGA-based SC execution engine that benefits from the inherent parallelism and high-speed processing capabilities of a hardware architecture. Synthesis results demonstrate a reduction in execution time of 61\% to 99\% for commonly used operation codes compared to CPU-based SC execution environments. Moreover, the execution time of Ethereum blocks on EVMx is up to 6$\times$ faster compared to analogous works in the literature. These results highlight the potential of the proposed architecture to accelerate SC execution and enhance the performance of EVM-compatible blockchains.
\end{abstract}

\begin{IEEEkeywords}
EVM, Ethereum, Blockchain, Smart Contracts, FPGA, Blockchain Hardware.
\end{IEEEkeywords}

\section{Introduction}

Ethereum blockchain pioneered \glspl{sc} and \glspl{dapp}, thereby transforming the blockchain ecosystem by enabling programmable, self-executing contracts and broadening the use cases of blockchain technology beyond transactions \cite{buterin2013ethereum}. Its ability to host programmable logic on-chain has revolutionized blockchain and created the foundation for the Web3 ecosystem \cite{buterin2013ethereum}. This is mainly enabled by its introduction of the Turing-complete \gls{evm}, which serves as the execution environment for \glspl{sc} \cite{song2024empirical}. Unlike Bitcoin, which is primarily a digital currency, Ethereum introduced a versatile platform where developers could build and deploy decentralized applications, tokenize assets, and create \gls{defi} ecosystems \cite{sriman2022decentralized}. 

Subsequently, many emerging blockchain platforms adopted the Ethereum model and became \gls{evm}-compatible. This compatibility enables developers to utilize the existing Ethereum ecosystem (e.g., programming languages, wallet software, plugins, etc.) to build and deploy applications across multiple blockchains, thereby enhancing scalability and alleviating congestion on the Ethereum network. Notable examples of \gls{evm}-compatible blockchains include Avalanche, Polkadot-Moonbeam, NEAR-Aurora, and Cardano \cite{jia2022evm}.

The \gls{evm} plays a central role in \gls{evm}-compatible blockchain platforms, as it is involved in multiple stages of transaction processing. For instance, when a new block is added to the blockchain, validators use the \gls{evm} to execute \glspl{sc} and verify transactions within the block \cite{grandjean2024ethereum}. Notably, 59\% of Ethereum transactions involve \gls{sc} execution, reflecting the extensive utilization of the \gls{evm}. Additionally, the remaining 41\% also use the \gls{evm} to check signatures, update account balances, and apply state changes \cite{jin2025chain, buterin2013ethereum}. Moreover, full and archival nodes, which maintain extensive histories of the blockchain, often depend on the \gls{evm} to synchronize and reconstruct a local copy of the blockchain by re-executing past transactions \cite{battah2021blockchain}. This synchronization process is often computationally intensive and can become a significant performance bottleneck \cite{cortes2024can}.

Several approaches have been proposed in the literature to improve the performance and efficiency of the \gls{evm} \cite{lu2020bpu, lu2023scu}. For instance, \cite{lu2020bpu} and \cite{lu2023scu} introduce a \gls{bpu} and a \gls{scu}, respectively—both are \gls{fpga}-based \gls{evm} implementations. These designs incorporate schedulers, decoders, and interpreters to evaluate and reorder \gls{sc} \glspl{op} for optimized execution, leveraging pipelining and parallelism to enhance performance. These works improve the execution time of \glspl{sc} compared to \gls{cpu}-based \glspl{evm}. 

However, \gls{sc} \glspl{op} are highly dependent \cite{contro2021ethersolve}, hence the cost of evaluating thousands of \glspl{op} for parallelism may outweigh the benefit of executing a few in parallel. To address this trade-off, our implementation bypasses the reordering stage entirely, opting instead to process operations in their original sequence while selectively leveraging opportunities for parallel execution during processing.

This work proposes EVMx, an \gls{fpga}-based \gls{evm} that offloads the execution of \glspl{sc} to a dedicated hardware device. The proposed architecture preserves the stack-based model of the existing \gls{evm}, ensuring seamless compatibility with existing \gls{evm}-based clients, while maintaining simplicity that prioritizes resource efficiency. Moreover, by leveraging the intrinsic speed that comes with the creation of dedicated hardware, the execution of \glspl{sc} is projected to be significantly faster compared to traditional \gls{sw}-based \glspl{evm}. Additionally, the dedicated hardware operates in isolation, removing \gls{sc} processing from the limitations of the general-purpose computer running the \gls{evm}. This enhances performance and introduces an extra layer of security by isolating the execution environment.

The remainder of this work is organized as follows. \autoref{sec:evm_overview} provides an overview of the \gls{evm}, outlining its key components and functionality. \autoref{sec:hardware_architecture} introduces the proposed hardware architecture of the \gls{evm}, offering a comprehensive description of its design and a detailed explanation of specific \gls{op} codes. Following this, \autoref{sec:results_discussions} presents the synthesis results of the proposed architecture and compares the execution times of various \gls{op} codes and blocks to highlight performance improvements. Finally, \autoref{sec:conclusion} summarizes the findings of this work and its future direction.

\section{EVM System Overview}
\label{sec:evm_overview}


Ethereum \glspl{sc} are primarily written in Solidity \cite{solidity} and compiled into \gls{evm} bytecode. This bytecode is a sequence of \glspl{op} deployed on the blockchain and executed by the \gls{evm}. Currently, the \gls{evm} supports 140 \glspl{op} \cite{opcodes}, each associated with a specific gas cost. Gas is a small fee that is associated with the execution of each \gls{op}. It is charged by the blockchain network infrastructure to the sender of the \gls{sc} to process a given transaction. The fee is measured in gwei (gigawei) and paid in \gls{eth}, Ethereum’s native currency. The gas fee model serves as a safeguard against \gls{dos} attacks by preventing excessive computation \cite{bistarelli2020ethereum}. The gas limit is the maximum amount of unit gas a user is willing to spend on a transaction. If the gas limit is exhausted, an $out-of-gas$ exception is thrown by the \gls{evm}, halting further processing and reverting any changes \cite{lu2023scu}.

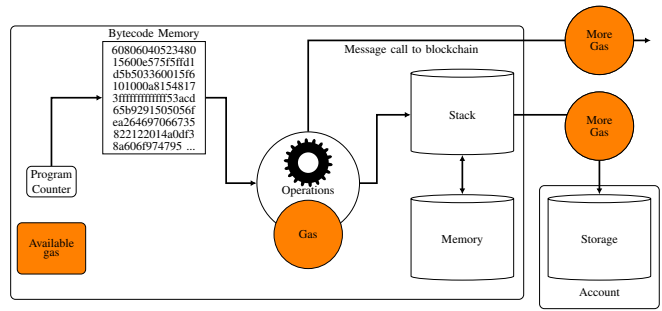
\begin{figure}[]
\centering
    \resizebox{0.482\textwidth}{!}{
    
    \begin{tikzpicture}
    \node [rectangle, rounded corners, minimum width=1cm, minimum height=0.5cm, draw](pc){\shortstack{Program\\Counter}};
    \node [fill=orange, rectangle, rounded corners, minimum width=2cm, minimum height=1.5cm, draw](gas) at ($(pc.south)+(0, -1.5)$){\shortstack{Available\\gas}};
    \node [rectangle, minimum width=3cm, minimum height=3cm, draw](evmcd) at ($(pc.north)+(3, 2)$){\shortstack{
                60806040523480\\
                15600e575f5ffd1\\
                d5b503360015f6\\
                101000a8154817\\
                3fffffffffffff53acd\\
                65b9291505056f\\
                ea264697066735\\
                822122014a0df3\\
                8a606f974795 ...}};
    \node at ([yshift=1.85cm]evmcd) {Bytecode Memory};
    \node [circle, minimum size=3cm, draw](c0) at ($(evmcd.east)+(3, -2.5)$) {}; 
    \begin{scope}[shift={($(c0)+(0,0.6)$)}, scale=0.3]
        \fill[even odd rule] \gear{18}{2}{2.4}{10}{2}{1};
    \end{scope}
    \node at ([yshift=-0.25cm]c0) {Operations};
    
    \node [fill=orange, circle, minimum size=2cm, draw](c1) at ($(evmcd.east)+(3, -4)$) {Gas}; 
    \node[cylinder, shape border rotate=90, draw,minimum height=2.5cm,minimum width=3cm](stack) at ($(c0.east)+(3,2)$){}; 
    \node at (stack) {Stack};
    \node[cylinder, shape border rotate=90, draw,minimum height=2.5cm,minimum width=3cm](memory) at ($(stack.south)+(0,-2.5)$){}; 
    \node at (memory) {Memory};
    \node[cylinder, shape border rotate=90, draw,minimum height=2.5cm,minimum width=3cm](storage) at ($(memory.east)+(2.5,0)$){}; 
    \node at (storage) {Storage};
    \node [fill=white, circle, minimum size=2cm, draw](c2) at ($(storage.north)+(0, 2)$) {Gas}; 
    \node [fill=white, circle, minimum size=2cm, draw](c3) at ($(c2.north)+(0, 1.5)$) {Gas}; 
    \node [rectangle, rounded corners, minimum width=3.5cm, minimum height=3.6cm, draw](acc) at ($(storage)+(0,-0.2)$){};
    \node at ([yshift=-1.3cm]acc) {Account};
    \node [rectangle, rounded corners, minimum width=15cm, minimum height=8cm, draw](rec) at ($(c0)+(-1.2,0.6)$){};

    \draw[line:thick](pc)|-(evmcd);
    \draw[line:thick](evmcd.east) -- ($(evmcd.east)+(0.6,0)$) |- (c0.west);
    \draw[line:thick](c0.north) |- ($(c3.east)+(0.5,0)$)node[xshift=-7cm, yshift=-0.3cm]{Message call to blockchain};
    \draw[line:thick](stack.east) -| (storage.north);
    \draw[line:thick](c0.east) -- ($(c0.east)+(0.5, 0)$) |- (stack.west);
    \draw[line:thickAr](stack.south) -- (memory.north);

    \node [fill=orange, circle, minimum size=2cm, draw](c2) at ($(storage.north)+(0, 2)$) {\shortstack{More\\Gas}}; 
    \node [fill=orange, circle, minimum size=2cm, draw](c3) at ($(c2.north)+(0, 1.5)$) {\shortstack{More\\Gas}}; 
\end{tikzpicture}
    
    }
    \caption{A simplified general structure of the \gls{evm}. Adapted from \cite{bistarelli2020ethereum}.}
   \label{fig:evm_diag} 
\end{figure}

\autoref{fig:evm_diag} illustrates the general structure of the \gls{evm}. The \gls{evm} follows a stack-based architecture and consists of several key components \cite{bistarelli2020ethereum}. These components are:  

\begin{enumerate}  
    \item \textbf{Stack:} A \gls{lifo} structure that holds up to 1024 words of 32 bytes each. It temporarily stores intermediate results during contract execution. The cost of stack operations is relatively low.  
    \item \textbf{Memory:} A dynamically allocated, temporary storage area used during execution to hold \gls{sc} data. Memory expansion increases costs, making its usage moderately expensive.  
    \item \textbf{Storage:} A persistent key-value store, where both keys and values are 32 bytes long. \Glspl{sc} use storage to maintain state variables across transactions. Due to its permanence, accessing storage is costly.  
    \item \textbf{Bytecode Memory:} A persistent part of a \gls{sc}’s account state, containing the bytecode executed by the \gls{evm}.  
    \item \textbf{Program Counter:} It tracks the current execution position within the \gls{sc} bytecode, determining which \gls{op} to execute next.  
    \item \textbf{Gas:} The total execution cost required to process a \gls{sc}. It is measured in gas units.
\end{enumerate}  


\section{Proposed Hardware Architecture of EVMx}
\label{sec:hardware_architecture}
\begin{figure*}[t]
\centering
    \resizebox{\textwidth}{!}{
    
    \begin{tikzpicture}
\node [rectangle, rounded corners, minimum width=23.5cm, minimum height=8cm, draw](boundary) at (9.5cm,-1.5cm){};
\node[rectangle, minimum width=1.3cm, minimum height=2.5cm, fill=yellow, draw](evm_code){\shortstack{BCM}};
\node[rectangle, minimum width=0.5cm, minimum height=0.5cm, fill=yellow, draw](pc) at ($(evm_code)+(-1.5, 0.5)$){PC};
\node[rectangle, minimum width=1cm, minimum height=1cm, draw](fsm) at ($(evm_code)+(1.5, -2)$){FSM};
\node[rectangle, minimum width=1cm, minimum height=1cm, fill=yellow, draw](gas) at ($(fsm)+(-1, -1.5)$){GS};
\node [rectangle, rounded corners, minimum width=0.6cm, minimum height=0.5cm, draw](pad0) at ($(evm_code)+(1.5, -0.5)$){\texttt{pad}};
\node [rectangle, minimum width=0.6cm, minimum height=1.5cm, draw](accm) at ($(evm_code)+(1.5, 0.7)$){r0};
\node[isosceles triangle, scale=0.6, draw] (clk00) at ($(accm)+(-0.22, -0.6)$){};
\node [trapezium, draw, rotate=-90, minimum width=1.5cm, minimum height=0.5cm](mux0) at ($(accm)+(1.5, -0.5)$){}; 
\node[rectangle, minimum width=1.3cm, minimum height=2.5cm, fill=yellow, draw](stack) at ($(mux0)+(1.5, -0.3)$){STK};
\node[rectangle, minimum width=0.6cm, minimum height=1.5cm, draw](reg0) at ($(stack)+(1.5, -1)$){r2};
\node[isosceles triangle, scale=0.6, draw] (clk0) at ($(reg0)+(-0.22, -0.6)$){};
\node[rectangle, minimum width=0.6cm, minimum height=1.5cm, draw](reg01) at ($(reg0)+(0, 2)$){r1};
\node[isosceles triangle, scale=0.6, draw] (clk0) at ($(reg01)+(-0.22, -0.6)$){};
\node[rectangle, minimum width=0.6cm, minimum height=1.5cm, draw](reg1) at ($(reg0)+(0, -2)$){r3};
\node[isosceles triangle, scale=0.6, draw] (clk1) at ($(reg1)+(-0.22, -0.6)$){};
\node [trapezium, draw, rotate=-90, scale=3](mux1) at ($(stack)+(3.3, 0.8)$){}; 
\node [trapezium, draw, rotate=-90, scale=2](mux2) at ($(mux1)+(-0.4, -1.5)$){}; 
\node[rectangle, minimum width=1cm, minimum height=1cm, draw](alu) at ($(reg1)+(1.5, -0.7)$){ALU};
\node[rectangle, minimum width=0.6cm, minimum height=1.5cm, draw](reg2) at ($(alu)+(1.2, -0.3)$){r4};
\node[isosceles triangle, scale=0.6, draw] (clk0) at ($(reg2)+(-0.22, -0.6)$){};
\node[rectangle, minimum width=1.3cm, minimum height=2.5cm, fill=yellow, draw](mem) at ($(mux1)+(1.8, -0.7)$){MEM};
\node[rectangle, minimum width=1.3cm, minimum height=2.5cm, fill=yellow, draw](strg) at ($(reg1)+(4.6, -0.7)$){STR};
\node[rectangle, minimum width=1.3cm, minimum height=2.5cm, draw](return) at ($(strg)+(1.8, 1)$){RTN};
\node[rectangle, minimum width=0.6cm, minimum height=1.5cm, draw](reg3) at ($(mem)+(2, 1)$){r5};
\node[isosceles triangle, scale=0.6, draw] (clk3) at ($(reg3)+(-0.22, -0.6)$){};
\node [trapezium, draw, rotate=-90, scale=3](mux3) at ($(reg3)+(2.8, 0)$){}; 
\node[rectangle, minimum width=1.3cm, minimum height=2cm, draw](kecc) at ($(mux3)+(2, 0)$){KEC};
\node[rectangle, minimum width=1cm, minimum height=1cm, draw](rpl) at ($(reg3)+(1, -1.4)$){RPL};
\node[rectangle, minimum width=1cm, minimum height=1cm, draw](addr) at ($(kecc)+(2, -0.2)$){ADR};
\node[rectangle, minimum width=0.6cm, minimum height=1.5cm, draw](reg4) at ($(addr)+(0, -1.4)$){r6};
\node[isosceles triangle, scale=0.6, draw] (clk4) at ($(reg4)+(-0.22, -0.6)$){};
\node[rectangle, minimum width=1cm, minimum height=1cm, draw](delta) at ($(clk4)+(2, -1)$){$\delta$};

\draw[line:thick](pc.east) -| ($(pc.east)+(0.2, 0)$) |- (evm_code.west);
\draw[line:thick](evm_code.east) -| ($(evm_code.east)+(0.1, 0)$) |- ($(fsm.west)+(0,0.2)$);
\draw[line:thick](evm_code.east) -| ($(evm_code.east)+(0.1, 0)$) |- (pad0.west);
\draw[line:thick](evm_code.east) -| ($(evm_code.east)+(0.1, 0)$) |- (accm.west);
\draw[line:thick]($(accm.east)+(0, 0.3)$) -| ($(accm.east)+(0.1, 0)$) |- (mux0.south);
\draw[line:thick](pad0) -| ($(pad0.east)+(0.1, 0)$) |- ($(mux0.south)+(0,-0.2)$);
\draw[line:thick](mux0.north) |- ($(stack.west)+(0, 0.3)$);
\draw[line:thick,red](stack.east) -- ($(stack.east)+(0.2,0)$) |- ($(reg1.west)+(0,0.2)$);
\draw[line:thick,red](stack.east) -- ($(stack.east)+(0.2,0)$) |- ($(reg0.west)+(0,0.4)$);
\draw[line:thick,red](stack.east) -- ($(stack.east)+(0.2,0)$) |- ($(reg01.west)+(0,0.2)$);
\draw[line:thick]($(reg01.east)+(0,0.3)$) -- ($(reg01.east)+(0.3,0.3)$) |- ($(mux2.south)+(0,-0.3)$);
\draw[line:thick]($(reg01.east)+(0,0.3)$) -- ($(reg01.east)+(0.3,0.3)$) |- ($(mux1.south)+(0,-0.4)$);
\draw[line:thick](mux1.north) -- ($(mux1.north)+(0.3,0)$) |- ($(mem.west)+(0,0.3)$);
\draw[line:thick](mux2.north) -- ($(mux2.north)+(0.3,0)$) |- node[above,xshift=0.2cm]{\texttt{sze}}($(mem.west)+(0,-0.3)$);
\draw[line:thick]($(reg1.east)+(0,0.3)$) -- ($(reg1.east)+(1.8,0.3)$) node[right,yshift=0.8cm]{\texttt{oft}}|- ($(mem.west)+(0,-0.8)$);
\draw[line:thick]($(reg1.east)+(0,0.3)$) -- ($(reg1.east)+(0.3,0.3)$) |- ($(alu.west)+(0,0.3)$);
\draw[line:thick,red](stack.east) -- ($(stack.east)+(0.2,0)$) |- ($(alu.west)+(0,-0.3)$);
\draw[line:thick]($(reg1.east)+(0,0.3)$) -- ($(reg1.east)+(1.8,0.3)$) |- ($(strg.west)+(0,0.8)$);
\draw[line:thick,red](stack.east) -- ($(stack.east)+(0.2,0)$) |- ($(strg.west)+(-0.3,-1.5)$) |- ($(strg.west)+(0,-0.3)$);
\draw[line:thick](alu.east) -- ($(alu.east)+(0.2,0)$) |- ($(reg2.west)+(0,0.3)$);
\draw[line:thick]($(mem.east)+(0,0.3)$) -- ($(mem.east)+(0.3,0.3)$) |- ($(reg3.west)+(0,0.3)$);
\draw[line:thick]($(reg3.east)+(0,0.4)$) |- ($(mux3.south)+(0,0.4)$);
\draw[line:thick](mux3.north) -- ($(mux3.north)+(0.3,0)$) |- ($(kecc.west)+(0,0.3)$);
\draw[line:thick]($(kecc.east)+(0,0.3)$) -- ($(kecc.east)+(0.3,0.3)$) |- (addr.west);
\draw[line:thick]($(kecc.east)+(0,0.3)$) -- ($(kecc.east)+(0.3,0.3)$) |- ($(reg4.west)+(0,0.3)$);
\draw[line:thick]($(reg4.east)+(0,0.3)$) -- ($(reg4.east)+(0.3,0.3)$)node[yshift=-0.4cm,right]{\texttt{d}} |- ($(delta.west)+(0,0.3)$);
\draw[line:thick]($(mem.east)+(0,0.3)$) -- ($(mem.east)+(0.3,0.3)$) |- ($(return.west)+(0,0.5)$);
\draw[line:thick]($(rpl.west)+(-0.5,0.3)$)node[xshift=-0.5cm]{\texttt{sAddr}} -- ($(rpl.west)+(0,0.3)$);
\draw[line:thick]($(rpl.west)+(-0.5,-0.3)$)node[xshift=-0.4cm]{\texttt{sNoc}} -- ($(rpl.west)+(0,-0.3)$);
\draw[line:thick](rpl.east) -- ($(rpl.east)+(0.2, 0)$) |- (mux3.south);
\draw[line:thick](delta.east) -| ($(delta.east)+(0.3,-0.7)$)node[xshift=-0.4cm, yshift=-0.2cm]{\texttt{k}} -| ($(mux3.south)+(-0.3,-0.4)$) -- ($(mux3.south)+(0,-0.4)$);
\draw[line:thick,red]($(delta.west)+(-0.7,-0.3)$)node[xshift=-0.6cm,black]{\texttt{oStack}} -- ($(delta.west)+(0,-0.3)$);
\draw[line:thick]($(kecc.east)+(0,0.3)$) -| ($(kecc.east)+(0.3,1.2)$) -| ($(mux0.south)+(-0.3, 0.6)$) |- ($(mux0.south)+(0, 0.6)$);
\draw[line:thick,red](stack.east) -- ($(stack.east)+(0.2,0)$) |- ($(pc.west)+(-0.3, 1.2)$) |- (pc.west);
\draw[line:thick](addr.east) -- ($(addr.east)+(0.3,0)$) |- ($(mux0.south)+(-0.4,2.1)$) |- ($(mux0.south)+(0,0.4)$);
\draw[line:thick]($(evm_code.west)+(-0.3,2.8)$)node[above]{\texttt{bytcd}} |- ($(evm_code.west)+(0,0.7)$);
\draw[line:thick]($(reg2.east)+(0,0.3)$) -| ($(reg2.east)+(0.2,-1)$) -| ($(mux0.south)+(-0.4,-0.4)$) |- ($(mux0.south)+(0,-0.4)$);
\draw[line:thick]($(reg0.east)+(0,0.3)$) -| ($(reg0.east)+(0.1,-1)$) -| ($(mux0.south)+(-0.3,-0.6)$) |- ($(mux0.south)+(0,-0.6)$);
\draw[line:thick]($(mem.east)+(0,0.3)$) -| ($(mem.east)+(0.3,1.6)$) -| ($(mux1.south)+(-0.3, 0.6)$) |- ($(mux1.south)+(0, 0.5)$);
\draw[line:thick,red](stack.east) -| ($(stack.east)+(0.2, 2)$) -| ($(mux1.south)+(-0.6,1)$) |- ($(mux1.south)+(0,-0.1)$);
\draw[line:thick,red](stack.east) -| ($(stack.east)+(0.2, 2)$) -| ($(mux1.south)+(-0.6,1)$) |- ($(mux2.south)+(0,0.3)$);
\draw[line:thick,blue]($(mux1.south)+(-0.45, 2)$)node[xshift=0.6cm,above,black]{\texttt{initData}} |- ($(mux1.south)+(0, 0.2)$);
\draw[line:thick,blue]($(mux1.south)+(-0.45, 2)$) |- ($(mux0.south)+(-0.56, 2.2)$) |- ($(mux0.south)+(0, 0.2)$);
\draw[line:thick]($(strg.east)+(0,1)$) -| ($(strg.east)+(0.3,-2)$)node[yshift=0.2cm,below]{\texttt{oStore}};
\draw[line:thick]($(return.east)+(0,1)$) -| ($(return.east)+(0.2,-3)$)node[yshift=0.2cm,below]{\texttt{retVal}};
\draw[line:thickAr](fsm.south) |- (gas.east);
\draw[line:thick]($(fsm.west)+(-3.4,-0.2)$)node[xshift=-0.5cm]{\texttt{start}} -- ($(fsm.west)+(0,-0.2)$);
\draw[line:thick]($(gas.west)+(-2.5,0)$)node[xshift=-0.4cm]{\texttt{gval}} -- (gas.west);
\draw[line:thick]($(reg01.east)+(0,0.3)$) -| ($(reg01.east)+(0.2,1.8)$)node[xshift=-0.3cm,above]{\texttt{val}};
\draw[line:thick]($(boundary.west)+(-0.5, -3)$)node[below]{\texttt{clk}} -- ($(boundary.west)+(0, -3)$);

\end{tikzpicture}

    }
    \caption{Block diagram of the proposed architecture of EVMx. The architecture receives the \gls{sc} bytecode via the \texttt{bytcd} interface and sequentially executes the instructions to fulfill the transaction request.}
    \label{fig:evm_arch} 
\end{figure*}
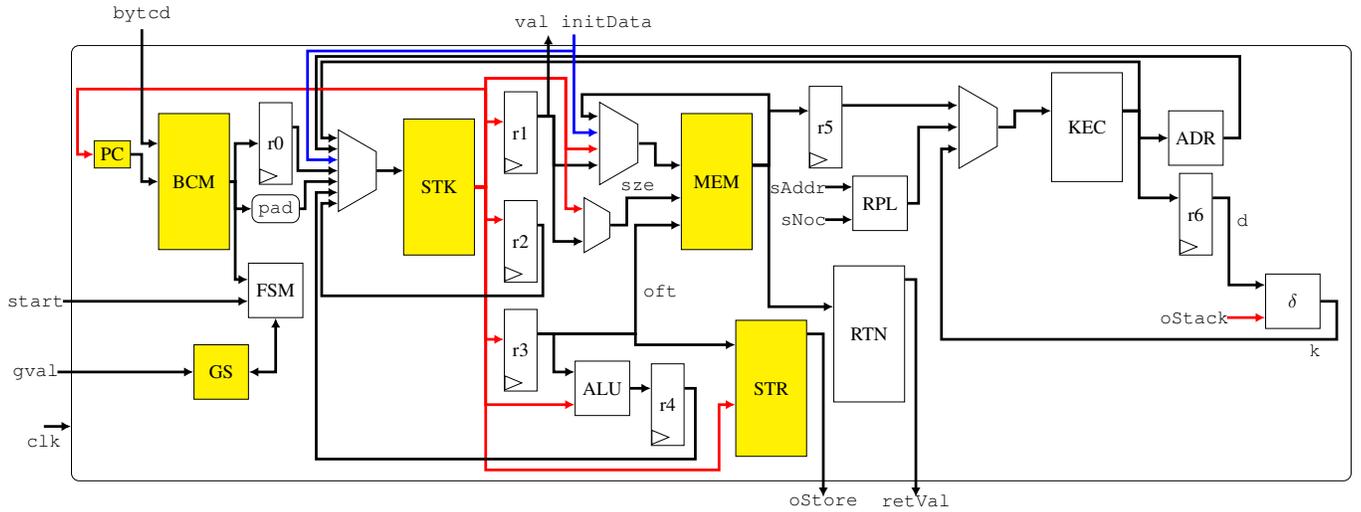
This section presents the proposed architecture of EVMx. Each component of the \gls{evm}, as outlined in \autoref{sec:evm_overview}, will be examined within the context of this architecture. Additionally, we will demonstrate the execution process of the CREATE2 \gls{op} to illustrate the system's functionality.

\autoref{fig:evm_arch} illustrates the proposed architecture of EVMx. The \gls{evm} components discussed in \autoref{sec:evm_overview} are highlighted in yellow and discussed below. 

\subsection{Stack}
The \gls{stk} processes 32-byte data during push-and-pop operations. To accommodate this, \texttt{pad0} pads the 1-byte data from \gls{bcd}. However, certain \gls{evm} operations, such as PUSH32, require reading multiple bytes from \gls{bcd} before pushing them to the stack. Therefore, r0 is a left-shift register that collects the data before pushing it. 

\subsection{Memory}
\Gls{mem} consists of 2\,768 words, each with 1 byte. Moreover, it is byte-addressable. Data can be written in chunks of either 1 or 32 bytes, but only 32 bytes can be read at a time. Additionally, many \gls{mem} \glspl{op} use an offset and a size parameter. The offset specifies the starting address in \gls{mem} for reading or writing, while the size defines the number of bytes to be copied or written. As a result, \gls{mem} contains two addresses, \texttt{oft} for offset and \texttt{sze} for the size.

\subsection{Storage}
The \gls{str} component has a depth of 1\,024, where the key parameter serves as the address and the value parameter represents the input. This approach inherently limits the number of key-value pairs that can be stored in \gls{str}. However, \glspl{sc} typically minimizes storage usage due to its high cost.

\subsection{Bytecode Memory and Gas}
The \gls{bcd} module stores the bytecode of the \gls{sc}. It can hold up to 32\,768 one-byte words of \glspl{op} and operands. When an \gls{op} is read from the \gls{bcd}, it is sent to the \gls{fsm}. The \gls{fsm} manages data flow by controlling read and write operations across components based on the executing \gls{op}. Moreover, the gas component (GS) is a counter loaded with the gas limit. The \gls{fsm} also ensures that the gas cost associated with the executing \gls{op} is deducted from the total gas.

\subsection{Program Counter}
Apart from incrementing sequentially from zero, the \gls{pc} can receive input from the \gls{stk} to support \glspl{op} like JUMP, which may require the \gls{pc} to move to a specific location. Also, the \gls{pc} outputs a 15-bit address that spans the entire \gls{bcd}, executing all the \glspl{op}.

\subsection{Arithmetic Logic Unit}
The \gls{alu} component performs various arithmetic and logical operations, including addition, multiplication, division, shifting, and modulo operations. However, multiplication, modulo operations, and division are the most computationally expensive operations. Therefore, we explore efficient implementation techniques to optimize performance.

To perform multiplication, we utilize the shift-and-add algorithm, which performs multiplication by shifting \cite{pathan2017optimised}. Moreover, we utilize a non-restoring algorithm that employs shifting to perform division and modulo operations \cite{jun2012modified}. These techniques avoid the use of resources such as \glspl{dsp} blocks, which impact the operating frequency and significantly increase the size of the device. Also, the algorithms handle all edge cases. For instance, when dividing by a power of two, the operation is reduced to a simple right shift, which is highly efficient in hardware.


\subsection{Other Components}
The \gls{kecc} component computes the Keccak256 hash digest, while the \gls{addr} component extracts the 20-byte Ethereum address from the digest. Additionally, the $\delta$ component calculates the output \texttt{k} using \eqref{eq:addr}.  

\begin{equation}
    \texttt{k} = \text{0xff} \| \texttt{sAddr} \| \texttt{salt} \| \texttt{d},
    \label{eq:addr}
\end{equation}

where $\|$ is the concatenation symbol, 0xff is the required byte prefix for the CREATE2 \gls{op}, \texttt{sAddr} is the address of the account deploying the new contract, \texttt{salt} is a user-defined input, and \texttt{d} is the Keccak256 digest. Also, the RPL component computes the \gls{rpl} encoding for executing the CREATE \gls{op}, where \texttt{sNoc} is the sender's nonce.

\subsection{\gls{evm} Inputs and Outputs}

The \gls{evm} has several inputs and outputs. The \texttt{initData} input is an array containing initialization data used at the start of execution. The \texttt{oStore} signal provides access to stored data, while \texttt{retVal} retrieves the returned data from \gls{retn} after a \gls{sc} is executed. Additionally, \texttt{val} represents the value sent to a newly created contract.  

The proposed architecture operates with a single synchronous clock (\texttt{clk}). The \texttt{start} signal initiates the \gls{sc} execution process, while \texttt{gval} loads the total available gas. Also, bytecode (\texttt{bytcd}) loads the bytecode of a compiled \gls{sc} to the \gls{bcd} component.

\subsection{Executing CREATE2 \Gls{op}}
The CREATE2 \gls{op} generates a new Ethereum account and associates it with a specific code at a predictable address using \eqref{eq:addr} \cite{opcodes}. The newly created address is then pushed onto the STACK. During execution, the CREATE2 \gls{op} pops four values from \gls{stk}: \texttt{value}, \texttt{offset}, \texttt{size}, and \texttt{salt}. \texttt{Value} specifies the amount of cryptocurrency (in wei) to be sent to the new account. The \texttt{offset} variable indicates the byte offset in \gls{mem} where the initialization code for the new account is stored, while \texttt{size} defines the length (in bytes) of the code to be copied. Finally, \texttt{salt} is a 32-byte user-defined value that ensures that a unique address is created.

In \autoref{fig:evm_arch}, when the \gls{fsm} receives the CREATE2 \gls{op}, it first pops \texttt{value} from the \gls{stk}. Next, it pops \texttt{offset} and stores the popped \texttt{value} in r1. It then pops \texttt{size} and stores the popped \texttt{offset} in r3 in parallel. The system then reads data from memory based on \texttt{offset} and \texttt{size}, accumulating it in the left-shift register r5. Once the data is fully copied, its digest is computed using \gls{kecc} and stored in r6 as output \texttt{d}.  

When r6 is loaded, \texttt{salt} is popped in parallel from the \gls{stk}. Then, $\delta$ computes \texttt{k} as shown in \eqref{eq:addr}, and the result is passed through the \gls{kecc} component once more. Finally, the first 20 bytes of the resulting Keccak256 digest are pushed onto the \gls{stk} as the address of the newly created account.

\subsection{Integrating EVMx with an Ethereum Client}

\gls{evm}-based blockchain networks rely on clients to connect with peers and maintain network operations. For instance, Ethereum users may run clients such as Geth \cite{geth}, Besu \cite{besu}, or Nethermind \cite{nethermind}. These clients are independent implementations of Ethereum that validate data according to its protocols, ensuring the integrity and security of the network. In Ethereum, each node typically runs two clients: a consensus client, responsible for implementing the consensus algorithm, and an execution client, which listens to new transactions broadcast within the network and executes them using the \gls{evm}. Even though the integration of EVMx with an \gls{evm}-compatible client is beyond the scope of this work, this section provides insights on how such an integration with an Ethereum consensus client may be implemented.

EVMx is a hardware accelerator, and the presented results were obtained using the \gls{pl} part of the ZCU104 \gls{fpga}. While there are multiple viable approaches to integrate the proposed accelerator with a software Ethereum client, one of them would consist of hosting the software Ethereum client on a Linux operating system executed on the \gls{ps} part of the \gls{fpga}. By writing an appropriate kernel module, the Ethereum client could communicate with EVMx, notably transferring the smart contract bytecode, initiating the execution, and reading back the results.

\subsection{Potential Operational Limitations}

One of the main challenges faced by EVMx arises from memory limitations when executing large \glspl{sc}. Specifically, the \gls{bcd} currently stores the entire bytecode of the deployed contract locally. For large \glspl{sc}, this bytecode could exceed the available on-chip memory of the \gls{fpga}. Additionally, transferring a large volume of data via \texttt{bytcd} may significantly increase the latency of \gls{sc} execution. 

To mitigate the memory limitation, the full bytecode could instead be stored in external memory, where segments would be streamed into a smaller buffer within the \gls{fpga}. This would provide efficient, temporary access during execution. 

\section{Results and Discussions}
\label{sec:results_discussions}
This section presents the synthesis results of the proposed architecture. Synthesis was done using Vivado 2024.2. The functionality of the \gls{evm} was verified by comparing its states with \gls{op} executions on Playground \cite{opcodes}  and Remix \cite{remix}.  

\autoref{table:evm_arch} compares the resource utilization of EVMx against works in the literature. Overall, the proposed implementation requires fewer resources. In particular, it uses the least number of \glspl{lut}, registers, and \glspl{dsp}, indicating its suitability for resource-constrained applications. This efficiency may be due to the absence of additional scheduler and decoder blocks, as well as dedicated extra configurable units (Adder, stack, memory, multipliers, etc.) used in \cite{lu2020bpu} and \cite{lu2023scu} to execute frequently used \glspl{sc} in parallel. Moreover, the proposed architecture uses only  13\% and 2\% of the available \glspl{lut} and registers on a  Zynq UltraScale (XCZU7EV) \gls{fpga}, indicating that there is more space available for additional functionality or optimization. Also, \autoref{table:evm_compon} summarizes the resources utilized by some of the main components in EVMx.

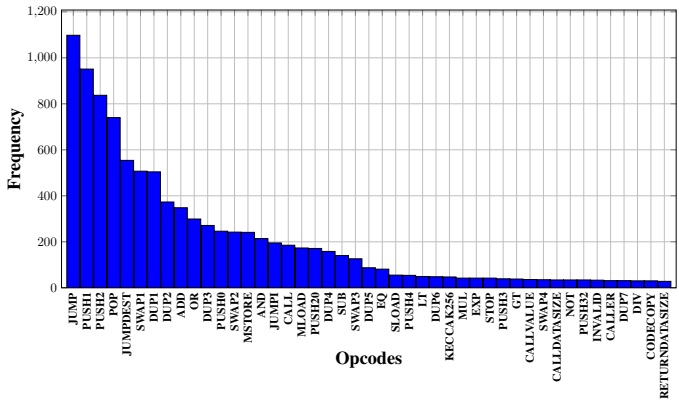
\begin{figure}[t]
    \centering
    \resizebox{0.5\textwidth}{!}{    
\begin{tikzpicture}
    \begin{axis}[
        xlabel={\bfseries Opcodes},
        ylabel={\bfseries Frequency},
        grid=both, 
        width=\textwidth, 
        height=0.5\textwidth, 
        scaled ticks=false, 
        ticklabel style={/pgf/number format/fixed}, 
        ybar, 
        bar width=10pt, 
        ymin=0, 
        symbolic x coords={JUMP, PUSH1, PUSH2, POP, JUMPDEST, SWAP1, DUP1, DUP2, ADD, OR, DUP3, PUSH0, SWAP2, MSTORE, AND, JUMPI, CALL, MLOAD, PUSH20, DUP4, SUB, SWAP3, DUP5, EQ, SLOAD, PUSH4, LT, DUP6, KECCAK256, MUL, EXP, STOP, PUSH3, GT, CALLVALUE, SWAP4, CALLDATASIZE, NOT, PUSH32, INVALID, CALLER, DUP7, DIV, CODECOPY, RETURNDATASIZE}, 
        xtick=data, 
        xtick align=center, 
        x tick label style={font=\small\bfseries, rotate=90, anchor=east}, 
        xtick align=inside, 
        enlarge x limits=0.02, 
        ylabel style={font=\Large\bfseries},
        xlabel style={yshift=-30pt, font=\Large\bfseries}, 
    ]
    \addplot[blue, fill=blue, draw=black] 
    table[
        col sep=comma, 
        x=x, 
        y=y  
    ] {opcodes1.csv};
    \end{axis}
\end{tikzpicture}

    }
    \caption{ Top 45 \glspl{op} used in verified Ethereum \glspl{sc} on Etherscan.} 
    \label{fig:opcodes}
\end{figure}

\autoref{fig:opcodes} presents the 45 most commonly used \glspl{op} in verified Ethereum \glspl{sc}, sampled from Etherscan \cite{etherscan}. As shown, certain \glspl{op}—such as JUMP, PUSH1, and PUSH2—appear frequently, while others are rarely used. Although EVMx is capable of executing the full set of \gls{evm} \glspl{op}, we focus our analysis on some of the most frequently used ones identified in \autoref{fig:opcodes} (e.g., PUSH, POP, and SWAP1), and evaluate their execution time on the proposed \gls{evm} architecture. Furthermore, we compare the execution times against those recorded by \gls{sw}-based \glspl{evm}.  

\begin{table}[t]
    \centering
    \caption{Resources utilized by crucial blocks in EVMx on a Zynq UltraScale \gls{fpga}.}
    \resizebox{\linewidth}{!}{
    \begin{threeparttable}
    \begin{tabular}{l c c c}
    \hline \hline
        Metrics &LUTs& Registers & RAM (kbits) \\ 
    \hline \hline
        Stack (STK)  & 10\,991  & 10  &   270 \\
        Memory (MEM)  & 2\,230  & 548  &   288 \\
        Storage (STR)  & 0  & 0  &   270 \\
        Bytecode Memory (BCM)  & 1\,097  & 22  &   522 \\
        Keccak (KEC)  & 2\,523  & 1\,620  &   0 \\
        Return Memory (RTN)  & 0 & 0  &   288 \\
        Arithmetic Logic unit (ALU)  & 5\,564  & 5\,295  &  0  \\
    \hline \hline
    \end{tabular}
    \end{threeparttable}
    }
    \label{table:evm_compon}
\end{table}

\begin{table}[t]
    \centering
    \caption{Comparison with similar implementations in the literature.}
    \resizebox{\linewidth}{!}{
    \begin{threeparttable}
    \begin{tabular}{l c c c}
    \hline \hline
        Metrics & BPU \cite{lu2020bpu} & SCU \cite{lu2023scu} & \textbf{EVMx} \\ 
    \hline \hline
        FPGA  & Zynq 7000  & Alveo U250  &   \textbf{Zynq XCZU7EV} \\
        Area \\
        ~~~kLUTs        &82.25 &142.82 & \textbf{29.80 (13\%)}    \\
        ~~~Registers    &67\,646 &98\,850 & \textbf{10\,720 (2\%)} \\
        ~~~DSP          &402   &225 & \textbf{0}             \\
        ~~~RAM (kbits)       &450    &3\,090 & \textbf{1\,638}           \\
        \hline
        Frequency (MHz) &100 &300 & \textbf{142.86}           \\
    \hline \hline
    \end{tabular}
    \end{threeparttable}
    }
    \label{table:evm_arch}
\end{table}

\autoref{table:comparison} presents a comparison of execution times for the selected \glspl{op} on various \glspl{cpu} against EVMx. To evaluate \gls{cpu} performance, we refer to the benchmarks in \cite{aldweesh2021opbench}, where three different Ethereum clients, PyEthApp, Go-Ethereum, and Parity, were executed on three \gls{cpu} platforms. These clients ensure compliance with Ethereum protocol rules and execute the \gls{evm}. The tested \glspl{cpu} were running Windows with an Intel i7 3.50 GHz processor, Linux with an Intel i7 3.50 GHz processor, and Mac with an Intel i5 2.8 GHz processor.  

Moreover, \autoref{table:comparison} shows the best-performing \gls{cpu} from each client in \cite{aldweesh2021opbench} and compares them against EVMx. Specifically, EVMx was evaluated against the PyEthApp client on Linux (LPy), the Go-Ethereum client on Windows (WGo), and the Parity client on Windows (WPa). The $\Delta$ column presents the relative percentage difference between EVMx and the minimum execution time recorded across the three \gls{cpu}-based platforms. It shows that across all the opcodes, EVMx outperforms its software counterparts. The execution-time reduction ranges from 61\% to 99\% where most reductions are above 95\%. This suggests that EVMx has a lower \gls{sc} execution time.

\begin{table}[t]
    \centering
    \caption{Processing time of \glspl{op} on \glspl{cpu} \cite{aldweesh2021opbench} and on EVMx}
    \resizebox{\linewidth}{!}{
    \begin{threeparttable}
       \begin{tabular}{l l l l l l l l l}\hline \hline
       Category &Opcode & Name  & Gas & LPy  &   WGo  &  WPa & \textbf{EVMx} & $\Delta$\tnote{a}\\
        & & & & ($n$s) & ($n$s) & ($n$s) & ($n$s) & \%\\
        \hline\hline
            \multirow{2}{*}{Arithmetic} & x01 & ADD              & 3 & 510   & 602  & 610  & \textbf{28} & 95\\
                                        & x03 & SUB              & 3 & 440   & 611  & 606  & \textbf{28} & 94\\ \hline   
            \multirow{3}{*}{Logic}      & x14 & EQ               & 3 & 430   & 571  & 604  & \textbf{28} & 93\\
                                        & x16 & AND              & 3 & 480   & 643  & 703  & \textbf{28} & 94\\
                                        & x17 & OR               & 3 & 490   & 646  & 701  & \textbf{28} & 94\\\hline
            \multirow{3}{*}{Environmental}  & x30 & ADDRESS      & 2 & 2770  & 1170  & 608 & \textbf{7} &  99\\
                                            & x33 & CALLER       & 2 & 3640  & 1142  & 614 & \textbf{7} &  99\\
                                            & x34 & CALLVALUE    & 2 & 80    & 556   & 604 & \textbf{7} &  91\\\hline
            \multirow{7}{*}{Memory/Stack}   & x50 & POP          & 2 & 220   & 570  & 605  & \textbf{7} &  97\\
                                            & x51 & MLOAD        & 3 & 6950  & 1838 & 666  & \textbf{259}& 61\\ 
                                            & x52 & MSTORE       & 3 & 2830  & 1726 & 684  & \textbf{245}& 64\\
                                            & x54 & SLOAD       &100 & 1990  & 694  & 701  & \textbf{21} & 97\\
                                            & x60 & PUSH1        & 3 & 260   & 600  & 640  & \textbf{14} & 95\\
                                            & x90 & SWAP1        & 3 & 310   & 528  & 550  & \textbf{28} & 91\\
                                            & x80 & DUP1         & 3 & 240   & 559  & 594  & \textbf{21} & 91\\
    \hline \hline
    \end{tabular}
    \begin{tablenotes}
            \item[a] $\Delta$=  $\frac{\text{Min(LPy, WGo, WPa)} - \text{EVMx}}{\text{Min(LPy, WGo, WPa)}}$ $\times$ 100, where Min is the minimum function. 
        \end{tablenotes}
    \end{threeparttable}}
    \label{table:comparison}
\end{table}

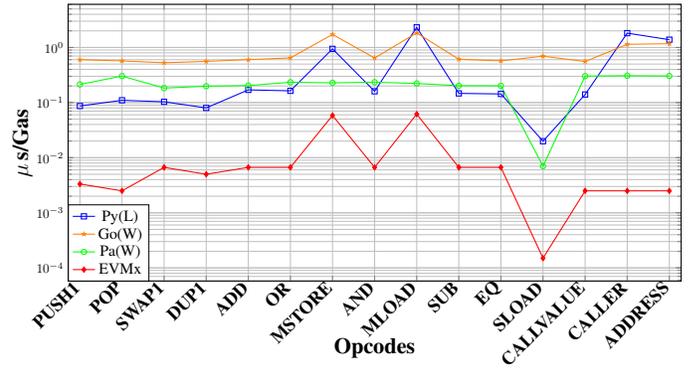
\begin{figure}[t]
    \centering
    \resizebox{0.5\textwidth}{!}{
    
    \begin{tikzpicture}
    \begin{axis}[
        xlabel={\bfseries Opcodes},
        ylabel={\bfseries $\mu$\,s/Gas},
        grid=both, 
        width=\textwidth, 
        height=0.5\textwidth, 
        legend style={font=\large,at={(0,0)}, anchor=south west}, 
        scaled ticks=false, 
        ticklabel style={/pgf/number format/fixed}, 
        symbolic x coords={PUSH1, POP, SWAP1, DUP1, ADD, OR, MSTORE, AND, MLOAD, SUB, EQ, SLOAD, CALLVALUE, CALLER, ADDRESS}, 
        xtick=data, 
        x tick label style={font=\Large\bfseries, rotate=45, anchor=east}, 
        y tick label style={font=\bfseries},
        xlabel style={yshift=-25pt, font=\LARGE\bfseries}, 
        ylabel style={font=\LARGE\bfseries}, 
        enlarge x limits=0.02,
        ymode=log, 
        ymajorgrids=true 
    ]

    \addplot[blue, thick, mark=square] 
    table[
        col sep=comma, 
        x=Opcode, 
        y=Py(L)
    ] {myfilen.csv};
    \addlegendentry{Py(L)}

    \addplot[orange, thick, mark=star] 
    table[
        col sep=comma, 
        x=Opcode, 
        y=Go(W)
    ] {myfilen.csv};
    \addlegendentry{Go(W)}

    \addplot[green, thick, mark=o] 
    table[
        col sep=comma, 
        x=Opcode, 
        y=Pa(W)
    ] {myfilen.csv};
    \addlegendentry{Pa(W)}

    \addplot[red, thick, mark=diamond*] 
    table[
        col sep=comma, 
        x=Opcode, 
        y=VHDL
    ] {myfilen.csv};
    \addlegendentry{EVMx}

    \end{axis}
\end{tikzpicture}

    }
    \caption{Execution time of selected \glspl{op} per unit gas.} 
    \label{fig:exTime_per_gas}
\end{figure}

Similar results are shown in \autoref{fig:exTime_per_gas}, which analyzes the execution time–to–gas ratio. The EVMx curve consistently lies below all other curves across all \glspl{op}, aligning with the $\Delta$ column in \autoref{table:comparison}. This demonstrates that EVMx achieves the lowest execution time per unit of gas, indicating that executing \glspl{op} on it is computationally less expensive than executing on \gls{cpu}-based \glspl{evm}. As a result, users consume fewer resources to perform the same tasks while benefiting from higher throughput.

 \autoref{table:block} compares the execution time of Ethereum blocks against previous works. Specifically, we analyze blocks \texttt{6653220}, \texttt{6653205}, and \texttt{6653209} which were also executed using a \gls{cpu}, \gls{bpu} \cite{lu2020bpu} and \gls{scu} \cite{lu2023scu}.  In \cite{lu2023scu}, multiple implementations of \gls{scu} (single and multi-core) were evaluated. For comparison, we use the best-performing \gls{scu} design and benchmark it against EVMx.  

\begin{table}[t]
    \centering
    \caption{Comparison of execution time of Ethereum blocks against works in literature.}
    \resizebox{1\linewidth}{!}{
    \begin{tabular}{lcccccc}\hline \hline
        Block & \# of  & CPU ($\mu s$)  & BPU ($\mu s$)  & SCU ($\mu s$) & \textbf{EVMx}  & $\Delta$\\
        &Txs& \cite{lu2023scu}&\cite{lu2020bpu}&\cite{lu2023scu}&($\mu s$)& $\%$\\
        \hline \hline
        6653205 & 16 & 3\,696 & 218.40 & 64.65 & \textbf{30.37} &53 \\
        6653220 & 9 & 4\,184 & 165.70 & 52.21 & \textbf{27.77} &47 \\
        6653209 & 159 & 46\,092 & 1\,595.80 & 372.01 & \textbf{232.82} & 37\\
        \hline \hline
    \end{tabular}}
    
    \label{table:block}
\end{table}

\autoref{table:block} shows that the execution time of all the blocks using EVMx is relatively close to each other; this is expected as the size of blocks should be fairly similar to avoid orphaned blocks \cite{jha2022study}. The table also shows that EVMx executes block \texttt{6653205} 122$\times$ faster than the \gls{cpu}, 7$\times$ faster than \gls{bpu}, and 2$\times$ faster than \gls{scu}. Moreover, EVMx executes block \texttt{6653220} 151$\times$ faster than the \gls{cpu}, 6$\times$ faster than \gls{bpu}, and 2$\times$ faster than \gls{scu}. Similarly, EVMx processes block \texttt{6653209} 198$\times$ faster than the \gls{cpu}, 7$\times$ faster than \gls{bpu}, and 2$\times$ faster than \gls{scu}. These results demonstrate EVMx's potential to improve the performance of the \gls{evm} in executing \glspl{sc}. The $\Delta$ column shows the difference in execution time between \gls{scu} and EVMx. 

\gls{scu} analyzes the bytecode to detect dependencies and reorders instructions accordingly before execution. Additionally, it uses a reordering buffer to write the results of out-of-order executions to storage elements correctly. These mechanisms may contribute to the time differences observed in \autoref{table:block}.



\section{Conclusion}
\label{sec:conclusion}

This work presents an \gls{fpga}-based \gls{evm} design that offloads smart contract execution to a dedicated \gls{hw} architecture. The design closely follows the conventional stack-based \gls{evm} model, maintaining simplicity while prioritizing resource efficiency. To achieve this, the proposed architecture integrates optimized modules that significantly reduce resource usage. Experimental results show a reduction of 61\% to 99\% in execution time for selected \glspl{op}. Additionally, the architecture executes entire Ethereum blocks up to 122$\times$ faster than \gls{cpu}-based \glspl{evm} and 6$\times$ faster than comparable \gls{fpga} implementations in the literature. Despite being preliminary, these results demonstrate the potential of EVMx to improve the execution time of the \gls{evm} compared to existing paradigms, consequently, improving the performance of \gls{evm}-compatible blockchain. Furthermore, the design occupies only 13\% of the available \glspl{lut} on a Zynq UltraScale \gls{fpga} and operates at a frequency of 142.86\,MHz, highlighting the potential for integrating additional functionality or applying further optimizations.

\bibliographystyle{IEEEtran}
\bibliography{IEEEabrv, ConfAbrv, references}

\end{document}